\documentclass[twocolumn,showpacs,preprintnumbers,amsmath,amssymb]{revtex4-1}
\usepackage{graphicx}
\usepackage{dcolumn} 
\usepackage{bm}      
\usepackage{xcolor}
\usepackage[mathscr,mathcal]{euscript}

\begin{document}

\title{Topological magnetic crystalline insulators and co-representation theory}
\author{Rui-Xing Zhang$^1$ and Chao-Xing Liu$^1$}
\affiliation{$^1$ Department of Physics, The Pennsylvania State University, University Park,
Pennsylvania 16802-6300}


\date{\today}

\begin{abstract}
	Gapless surface states of time reversal invariant topological insulators are protected by the anti-unitary nature of the time reversal operation. Very recently, this idea was generalized to magnetic structures, in which time reversal symmetry is explicitly broken, but there is still an anti-unitary symmetry operation combining time reversal symmetry and crystalline symmetry. These topological phases in magnetic structures are dubbed ``topological magnetic crystalline insulators''. In this work, we present a general theory of topological magnetic crystalline insulators in different types of magnetic crystals based on the co-representation theory of magnetic crystalline symmetry groups. We construct two concrete tight-binding models of topological magnetic crystalline insulators, the $\hat{C}_4\Theta$ model and the $\hat{\bf \tau}\Theta$ model, in which topological surface states and topological invariants are calculated explicitly. Moreover, we check different types of anti-unitary operators in magnetic systems and find that the systems with $\hat{C}_4\Theta$, $\hat{C}_6\Theta$ and $\hat{\bf \tau}\Theta$ symmetry are able to protect gapless surface states. Our work will pave the way to search for topological magnetic crystalline insulators in realistic magnetic materials.
\end{abstract}

 \maketitle
\section{Introduction}
In condensed matter physics, a topological state is a new type of quantum state of matter that cannot be classified by symmetry principles \cite{qi2011,hasan2010,moore2010,qi2010phystoday}. Topological phases of free fermions are usually characterized by an insulating bulk and metallic edge/surface states. The quantum Hall state is the first example of a topological state in two dimensions \cite{klitzing1980}. In a quantum Hall state, a gapless one-dimensional (1D) chiral edge state propagates along one direction at the 1D edge of a finite two-dimensional (2D) sample. Recently, it was realized that under certain symmetries, new topological phases can appear. For example, time reversal (TR) invariant topological insulators (TIs) were predicted in a system with TR symmetry in both two and three dimensions \cite{kane2005a,kane2005b,bernevig2006a,bernevig2006c,fu2007b,fu2007a,moore2007}. TR invariant TIs have been confirmed in various materials, such as HgTe quantum wells\cite{bernevig2006c,koenig2007}, BiSb\cite{hsieh2008}, Bi$_2$Se$_3$ family of materials\cite{hjzhang2009,xia2009,chen2009}, etc\cite{yan2012}, by different experimental methods, including angular-resolved photon emission spectroscopy (ARPES) \cite{hsieh2008,hsieh2009,xia2009,chen2009}, scanning tunnelling microscopy (STM) \cite{roushan2009,tzhang2009,alpichshev2010} and transport measurements\cite{koenig2007,brune2011b,qu2010,analytis2010}. The metallic edge/surface states of TR invariant TIs consist of two branches with opposite spin-polarization counterpropagating at a given edge/surface, and are thus dubbed ``helical edge/surface states''\cite{wu2006,xu2006}. Besides TR symmetry, topological states can also be protected by other types of symmetries, such as crystalline symmetry\cite{fu2011a,slager2012,fang2012a}, particle-hole symmetry\cite{qi2009b,lutchyn2010,fu2008}, etc. For example, SnTe systems have been theoretically predicted \cite{hsieh2012} and experimentally confirmed \cite{dziawa2012,tanaka2012,xu2012a} to be a topological crystalline insulator protected by mirror symmetry. It is believed that backscatterings are significantly suppressed for topological edge/surface states, so electric currents can flow with low dissipation in topological phases. Therefore, the research on topological phases has a potential application in electronics with low dissipation.

The gapless nature of helical edge/surface states in TR invariant TIs originates from double degeneracy due to Kramers' theorem\cite{kane2005a,kane2005b,xu2006,fu2007b}. Unlike unitary crystalline symmetry operators, TR symmetry operator $\hat{\Theta}$ is anti-unitary. For a TR invariant system, the TR operator sastifies $\hat\Theta^2=-1$ for spinful fermions and $\hat\Theta^2=1$ for spinless fermions or bosons. Kramers' theorem \cite{dresselhaus2008} tells us that when $\hat\Theta^2=-1$, a double degeneracy appears generally for each energy level. In a TR invariant TI, two branches of helical surface/edge states are related to each other by TR symmetry, forming the so-called ``Kramers' pairs''. For a generic momentum, one state of the ``Kramers' pairs'' at ${\bf k}$ is degenerate with the other at $-{\bf k}$. There are some special momenta satisfying the condition ${\bf k}=-{\bf k}+{\bf G}$ (${\bf G}$ is the reciprocal lattice vector), known as ``TR invariant momenta (TRIM)''. All the states at TRIM must be doubly degenerate. The Dirac point of the helical surface/edge states in TR invariant TIs is protected by this degeneracy. When TR symmetry is broken, the degeneracy at TRIM no longer exists, and the topological surface/edge states can be gapped, e.g. in the case of magnetically doped TIs \cite{chen2010c}.

From this discussion, we can see that the degeneracy induced by anti-unitary operators is essential for topologically non-trivial surface states. However, anti-unitary operators not only appear in a TR invariant system, but also exist in TR breaking systems, including various magnetic materials. For example, an anti-ferromagnetic system usually possesses an anti-unitary operator that combines the TR symmetry operator $\hat{\Theta}$ with a translation operator $\hat{\bf \tau}$. It has been shown that the anti-unitary operator $\hat{\bf \tau}\hat\Theta $ can also lead to a $Z_2$ topological phase, dubbed ``anti-ferromagnetic topological insulator'' \cite{mong2010,fang2013,liu2013c}. In the study of crystallography, the symmetry groups that include anti-unitary operators are known as magnetic symmetry groups, as an extension of the conventional crystalline symmetry groups. Just like the classification of crystal structures according to crystalline symmetry groups, different types of magnetic structures are classified by magnetic symmetry groups. Therefore, it is natural to ask what types of magnetic symmetry groups, as well as their corresponding magnetic structures, can support topologically non-trivial phases.

To answer this question, we develop a theory for topological phases in magnetic crystals, dubbed ``topological magnetic crystalline insulators (TMCIs)'', based on the ``co-representation'' theory of magnetic groups in this paper \cite{bradley1968,dresselhaus2008}. It is well-known that the degeneracy of energy states is related to irreducible representations of a crystalline symmetry group in the representation theory. However, the conventional representation theory can not be directly applied to magnetic groups due to the existence of anti-unitary operators. For example, in a TR invariant system, the double degeneracy due to Kramers' theorem of TR symmetry is regarded as an ``additional degeneracy'' that is not included in the conventional representation theory \cite{dresselhaus2008}. In order to understand this ``additional degeneracy'', Wigner first developed the so-called ``co-representation'' theory \cite{wigner1959,bradley1968}. The co-representation is different from the conventional representation because the multiplication rule is modified due to anti-unitary operators. Consequently, some useful concepts in the conventional representation theory, such as characters, can no longer be applied to the co-representation theory. The additional degeneracy of energy states due to anti-unitary operators can be described well in the co-representation theory. Since the $Z_2$ topological phases are closely related to the degeneracies induced by anti-unitary operators, the co-representation provides a natural approach to investigate $Z_2$ topological phases in magnetic structures.

In this paper, we will first review magnetic symmetry groups, and then discuss our approach to generalize $Z_2$ topological phases to a magnetic system based on the co-representation theory of magnetic symmetry groups. We will consider two concrete tight-binding models for TMCIs. One is related to the combination of TR symmetry and four-fold rotation symmetry, dubbed ``$\hat{C}_4\hat{\Theta}$ model'', while the other is related to the combination of TR symmetry and translation symmetry, dubbed ``$\hat{\bf \tau}\hat\Theta$ model''. We utilize the Wilson loop technique \cite{yu2011,kingsmith1993} to calculate the corresponding $Z_2$ topological invariant in these two models. Finally, we will generalize our discussion to other types of magnetic symmetry groups and show that a $Z_2\times Z_2$ topological magnetic crystalline phase exists in a system with $\hat{C}_6\hat{\Theta}$ symmetry. The paper is organized as follows: In Sec. II, we will summarize the co-representation theory of magnetic crystalline groups and discuss the principle used to define $Z_2$ topological invariants in magnetic structures. In Sec. III, we discuss two concrete models for $Z_2$ TMCIs, as well as the generalization to other magnetic symmetry groups. Conclusion is drawn in Sec. IV.

\section{Magnetic crystalline symmetry}

\subsection{Magnetic point groups and magnetic space groups}
We start from a review of the structure of magnetic symmetry groups and the so-called ``co-representation'' theory \cite{bradley1968,dresselhaus2008}. Magnetic symmetry groups include both unitary operations of crystalline symmetry and anti-unitary symmetry operations due to TR symmetry. Similar to point groups and space groups, we also have magnetic point symmetry groups and magnetic space symmetry groups. There are in total 122 magnetic point symmetry groups that can be classified into three types. Type I groups are just the ordinary point groups, denoted as ${\bf G}$, so there are 32 of them. Type II groups are the direct product of an ordinary point group ${\bf G}$ with the group $\{E,\hat{\Theta}\}$, where $E$ is the identity and $\hat{\Theta}$ is the TR symmetry. There are also 32 different types of type II groups. The most interesting groups are the type III groups, which take the form ${\bf M}={\bf G}+A{\bf G}$ where $A=\Theta R$ and the crystalline symmetry operation $R$ does not belong to ${\bf G}$. There are 58 type III magnetic point symmetry groups in total. For a type III group, half of the elements are unitary and the other half are anti-unitary. Magnetic space groups are constructed by the product of the translation symmetry group and the magnetic point symmetry group. 1651 magnetic space groups can also be categorized into three classes, 230 ordinary crystallographic space groups (Type I), 320 type II and 1191 type III groups. We can further divide the type III magnetic symmetry group into two classes: type IIIa where $R$ cannot be chosen to be a pure translation and type IIIb where $R$ is just a pure translation. For type IIIb, $R^2$ must be a translation operator, so in this case the subgroup ${\bf S}={\bf T}+\hat{\Theta} R{\bf T}$ characterizes the so-called magnetic Bravais lattice, where ${\bf T}$ is the translation symmetry group. There are 674 of type IIIa and 517 of type IIIb.

Before going into a detailed discussion, we will first illustrate our notations. We usually denote a magnetic symmetry group by ${\bf M}({\bf G})$ \cite{bradley1968}, where ${\bf M}$ is for the magnetic group and ${\bf G}$ is for its unitary part. The translation subgroup is denoted as ${\bf T}$. Below, we will use ${\bf t,w,v},\cdot\cdot\cdot$ for translation operators, $R,S,W,\cdot\cdot\cdot$ for unitary point symmetry operators and $A,B,\cdot\cdot\cdot$ for anti-unitary operators. TR symmetry is always denoted by $\hat{\Theta}$. An element in a space group is denoted as $\{S|{\bf t}\}$ where $S$ is for the point group symmetry and ${\bf t}$ is for the translation operation.

\subsection{Co-representation theory and degeneracy}
It is well-known that electronic states can serve as the basis for the construction of representations of a symmetry group and degeneracies in electronic band structures are directly determined by irreducible representations of the corresponding symmetry group \cite{dresselhaus2008}. The irreducible representations can be read out from the character table of a symmetry group. However, type II and type III magnetic symmetry groups contain anti-unitary operators, which are either a TR operation or a combination of a TR operation and another unitary operation. This prevents the construction of conventional representations. To find an appropriate description of a magnetic symmetry group ${\bf M}$, we may first start with the unitary part ${\bf G}$. Similar to the conventional case, we can act symmetry operators in ${\bf G}$ on the eigen states $|\psi_j\rangle$ $j=1,\cdots,d$ of a system. This leads to a $d$-dimensional representation, denoted as $\Delta$, of the unitary group ${\bf G}$. For the whole magnetic symmetry group ${\bf M}$, we take $|\psi_j\rangle$ and $A|\psi_j\rangle$ as basis where $A$ is an anti-unitary operator in ${\bf M}$. As shown in details in the appendix, the obtained matrices cannot form conventional representations due to the anti-unitary operators. Thus, they are dubbed ``co-representations'', denoted as $D$ below. There are two main differences between co-representations and conventional representations. Firstly, the multiplication rules are modified to (\ref{eq:MSG_DBS1})$\sim$(\ref{eq:MSG_DBC1}) for a co-representation in the appendix. Secondly, the character (the trace of a representation matrix) can be changed by an unitary transformation of the basis in the co-representation theory. Consquently, the character table does not have any meaning. Therefore, we can not apply the conventional method to determine the reducibility of a co-representation by its character table. Since the relation between the degeneracy and an irreducible co-representation still exists, the central goal of the co-representation theory is to develop a method to analyze the reducibility of a co-representation of a magnetic symmetry group. This is achieved by a method called the ``Herring rule'' \cite{herring1937a}, which will be described below. In this section, we would like to discuss some simple but most useful cases and show how the Herring rule works.

Let us focus on the case that $A=\hat{\Theta} R$ is the only generator in a magnetic symmetry group ${\bf M}$. Since TR symmetry operator $\hat{\Theta}$ commutes with any crystalline symmetry operator $R$, the magnetic symmetry group ${\bf M}$ and its unitary subgroup ${\bf G}$ are Abelian. Consequently, the unitary subgroup ${\bf G}$ only possesses one-dimensional (1D) irreducible representations. For a wave function $|\psi\rangle$ that forms a basis for the 1D representation $\Delta$ of the group ${\bf G}$, the matrix element $\langle\psi|A|\psi\rangle$ is given by
\begin{eqnarray}
	\langle\psi|A|\psi\rangle=\langle A^2\psi|A\psi\rangle=\Delta^*(A^2)\langle\psi|A|\psi\rangle
	\label{eq:MSG_matrixelement}
\end{eqnarray}
in which $A^2\in {\bf G}$. Therefore, if $\Delta^*(A^2)\neq 1$, $\langle\psi|A|\psi\rangle$ must equal to zero, implying that $|\psi\rangle$ and $A|\psi\rangle$ are orthogonal to each other. This argument indicates that the reducibility of the co-representation of ${\bf M}$ is determined by the representation $\Delta$ of the unitary group ${\bf G}$.

Now let us check the reducibility of the co-representation $D$ directly. The matrix forms of the co-representation $D$ are constructed as Eq. (\ref{eq:MSG_DS1}) and (\ref{eq:MSG_DB1}), which can be simplified in our Abelian case. Since any $S\in {\bf G}$ commutes with $A$, Eq. (\ref{eq:MSG_DS1}) is simplified as
\begin{equation}
	D(S)=\left(
	\begin{array}{cc}
		\Delta(S)&0\\0&\Delta^*(S)
	\end{array}
	\right).
	\label{eq:MSG_DS2}
\end{equation}
For any $B=AS$, Eq. (\ref{eq:MSG_DB1}) takes the form
\begin{eqnarray}
	&&D(B)=\left(
	\begin{array}{cc}
		0&\Delta(A^2)\Delta(S)\\
		\Delta^*(S)&0
	\end{array}	\right).
	\label{eq:MSG_DB2}
\end{eqnarray}
Next we need to check the condition when both the matrices (\ref{eq:MSG_DS2}) and (\ref{eq:MSG_DB2}) can be diagonalized by an unitary transformation $U$ simultaneously. There are three different cases. If $\Delta(S)$ is complex for some $S\in{\bf G}$ (case c), to keep $D(S)$ in Eq. (\ref{eq:MSG_DS2}) diagonal, we require the transformation matrix $U$ to be diagonal. However, any diagonal unitary tranformation matrix $U$ cannot diagonalize $D(B)$ in Eq. (\ref{eq:MSG_DB2}). So the 2D co-representation is irreducible in the case c.

For a real representation $\Delta$, $\Delta(S)=\Delta^*(S)$ can only take the values of $\pm 1$, so $\Delta(A^2)=\pm 1$ ($A^2\in{\bf G}$). Eq. (\ref{eq:MSG_DS2}) is an identity matrix, and will not be changed by any unitary transformation $U$. Thus, we only need to find a transformation matrix $U$ to diagonalize Eq. (\ref{eq:MSG_DB2}). For $\Delta(A^2)=1$ (case a), one can take the unitary transformation
\begin{equation}
	U=\frac{1}{\sqrt{2}}\left(
	\begin{array}{cc}
		1&1\\-1&1
	\end{array}
	\right)
\end{equation}
and $D(B)$ is transformed as
\begin{eqnarray}
	&&D'(B)=U^{-1}D(B)U^*=\left(
	\begin{array}{cc}
		-\Delta(S)&0\\0&\Delta(S)
	\end{array}
	\right)\nonumber\\
	&&B=AS\in A{\bf G}.
	\label{eq:MSG_transformation2}
\end{eqnarray}
Therefore, the co-representation is reducible in the case a.

For $\Delta(A^2)=-1$ (case b), $D(B)=-i\Delta(S)\sigma_y$ for $B=AS$ where $\vec{\sigma}$ denote Pauli matrices. Any unitary transformation matrix $U$ can be expanded as $U=e^{i\vec{\sigma}\cdot\vec{\theta}}$. Based on Eq. (\ref{eq:MSG_transformation}), direct calculation gives
\begin{eqnarray}
	&&D'(B)=e^{-i\sigma_{j}}(-i\Delta(S)\sigma_y)e^{-i\sigma_{j}}=e^{-i\sigma_{j}}e^{i\sigma_{j}}(-i\Delta(S)\sigma_y)\nonumber\\
	&&=-i\Delta(S)\sigma_y=D(B)\qquad j=x,z
\end{eqnarray}
and
\begin{eqnarray}
	&&D'(B)=e^{-i\sigma_{y}}(-i\Delta(S)\sigma_y)e^{i\sigma_{y}}\nonumber\\
	&&=-i\Delta(S)\sigma_y=D(B).
\end{eqnarray}
Therefore, any unitary transformation $U$ leaves the representation matrix $D(B)$ unchanged. We conclude that the co-representation is also irreducible in the case b.

In summary, according to the representation $\Delta$ of the unitary part ${\bf G}$, the co-representation of a magnetic point symmetry group ${\bf M}$ is reducible for the case a and irreducible for the case b and c. Although the above discussion is based on a simple situation with only one generator, a simple rule to determine the reducibility of a co-representation exists, which was first described by Herring \cite{herring1937a}. One needs to calculate the summation of the characters $\chi(B^2)$ for all $B\in A{\bf G}$ ($B^2\in {\bf G}$) and the obtained value determines which case the co-representation belongs to. The Herring rule can be summarized as
\begin{eqnarray}
	\sum_{B\in A{\bf G}} \chi(B^2)=\left\{
	\begin{array}{c}
		|G|\qquad \mbox{in case a},\\
		-|G|\qquad \mbox{in case b},\\
		0\qquad \mbox{in case c}.
	\end{array}
	\right.
	\label{eq:MSG_Herring1}
\end{eqnarray}
where $\chi$ is the character of the representation $\Delta$, $|G|$ is the element number of the group ${\bf G}$. As an example, we check this rule for our simple case,
\begin{eqnarray}
	\sum_{B\in A{\bf G}} \chi(B^2)=\sum_{S\in {\bf G}}\chi(A^2S^2)=\chi(A^2)\sum_{S\in {\bf G}}\chi^2(S).
\end{eqnarray}
Since ${\bf G}$ is a cyclic group, 1D representation can only take the form of $e^{i2\pi \frac{n}{N}}$ where $N$ should be a factor of $|G|$ ($|G|=MN$ where $N$ and $M$ are integers) and $n=1,\ldots,N$. The summation $\sum_{S\in {\bf G}}\chi^2(S)=M\sum_{n=1,\cdots,N}e^{i4\pi \frac{n}{N}}$ has the following possible values. If $N>2$, $\sum_{B\in A{\bf G}} \chi(B^2)=0$ corresponds to the case c of complex representations. If $N=1,2$, all the representations are real and the character $\chi(S)$ can only be $\pm 1$ for any $S\in {\bf G}$, so $\sum_{S\in {\bf G}}\chi^2(S)=NM=|G|$. Correspondingly, the summation $\sum_{B\in A{\bf G}} \chi(B^2)=\chi(A^2)|G|$ is determined by the sign of $\chi(A^2)$, which just corresponds to the case a or b. Therefore, the Herring rule (\ref{eq:MSG_Herring1}) does distinguish three different cases.

%

\subsection{Magnetic space group and $Z_2$ topological invariants}
Next we will illustrate our approach to determine the existence of $Z_2$ topological phases for a magnetic crystal with a given magnetic symmetry group. Our goal is to construct $Z_2$ topological invariants, in analogy to that defined for TR invariant TIs. From the examples of TR invariant TIs, $Z_2$ topological invariants are defined at TRIM in the Brillouin zone (BZ). Therefore, in this section, we need first to apply the Herring rule to understanding degeneracies in the BZ of a magnetic crystal with a magnetic space symmetry group.

An element in a space group can be represented by $\left\{ S|{\bf t+v} \right\}$ where $S$ is a point group operation, ${\bf t}$ is a lattice translation operation and ${\bf v}$ is zero or a non-primitive translation operation. For a magnetic space symmetry group ${\bf M}({\bf G})$, besides the conventional operations, we have additional anti-unitary operators with the form $A\left\{ S|{\bf t+v} \right\}$, where $A=\hat{\Theta}\left\{ R|{\bf w} \right\}$. Here $\left\{ R|{\bf w} \right\}$ is not a symmetry operation in the unitary part ${\bf G}$. For a space group, all the translation operations form an Ablien subgroup, which is usually described by the 1D representation $e^{i{\bf k}\cdot{\bf t}}$. Here ${\bf k}$ is the wave vector that labels the 1D representation. A point symmetry operation usually changes one momentum ${\bf k}$ to another ${\bf k'}$. All the symmetry operations, that preserve the momentum ${\bf k}$, form a subgroup of the whole space group. This subgroup is known as a wave vector group or a little group, denoted as ${\bf M_k}$ for a magnetic space symmetry group and ${\bf G_k}$ for its unitary part. The symmetry operations in a wave vector group are different for different ${\bf k}$, and it is more convenient to discuss the representation of a wave vector group. For an element $\{S|{\bf t+v}\}$ in a wave vector group at ${\bf k}$, the representation takes the form $\Delta_{\bf k}(\{S|{\bf t+v}\})=e^{i{\bf k\cdot(t+v)}}\Delta(S)$, where $\Delta(S)$ denotes the representation for the point symmetry part.

For a generic ${\bf k}$, the wave vector group may only contain translation symmetry operations. But at high symmetry momenta, point symmetry operations and anti-unitary operations can also exist. Here we are only interested in the anti-unitary operation $A=R\hat{\Theta}$. We denote the momenta ${\bf k}$ which is invariant under the operation $A$ as ${\bf K}_{\alpha}$. ${\bf K}_{\alpha}$ satisfies the condition
\begin{equation}
	R{\bf K}_{\alpha}=-{\bf K}_{\alpha}+{\bf g},
	\label{eq:MSG_Kinvariant}
\end{equation}
where $\alpha=1,2,\cdots$ denotes the number of the invariant momenta and ${\bf g}$ is a reciprocal lattice vector. These momenta are dubbed ``$A$-invariant momenta'' below. The $Z_2$ topological invariant can be defined on these $A$-invariant momenta ${\bf K}_{\alpha}$ in magnetic systems.

For the case of TR invariant TIs, one introduces an anti-symmetric matrix $w_{nm}({\bf K}_\alpha)=\langle \psi_{n,{\bf K}_\alpha}|\hat{\Theta}|\psi_{m,{\bf K}_\alpha}\rangle$ at TRIM ${\bf K}_\alpha$. The $Z_2$ topological invariant is defined as\cite{fu2006}
\begin{equation}
(-1)^{\Delta}=\prod_{\alpha}\frac{\sqrt{\det[w({\bf K}_\alpha)]}}{Pf[w({\bf K}_\alpha)]},
\end{equation}
where $\alpha=1,2,3,4$. "det" and "Pf" are determinant and Pfaffian of the anti-symmetric matrix $w$.
It has been shown that one $Z_2$ topological invariant can be defined at four TRIM in one plane. In two dimensions, there are four TRIM, so only one $Z_2$ invariant is possible. In three dimensions, there are eight TRIM in total and as a result, one can define four $Z_2$ invariants (one strong $Z_2$ invariant and three weak $Z_2$ invariants).

For magnetic systems, the anti-unitary operator $A$ plays the role of TR symmetry operator $\hat{\Theta}$ in TR invariant TIs. But there are two main differences. (1) Instead of TRIM, we require at least four $A$-invariant momenta that satisfy the condition (\ref{eq:MSG_Kinvariant}), in order to define a $Z_2$ topological invariant. (2) From the Herring rule, we know that double degeneracy only occurs when the states belong to the co-representation of the case b and c. Only in these cases, one can define an anti-symmetric matrix ${\cal A}_{nm}({\bf K}_\alpha)=\langle \psi_{n,{\bf K}_\alpha}|A|\psi_{m,{\bf K}_\alpha}\rangle$, where $|\psi_{n,{\bf K}_\alpha}\rangle$ is the eigen-state at invariant momenta ${\bf K}_\alpha$. By defining $\delta_{{\bf K}_\alpha}=\frac{\sqrt{\det{{\cal A}({\bf K}_\alpha)}}}{Pf[{\cal A}({\bf K}_{\alpha})]}$, the corresponding $Z_2$ topological invariant for TMCIs is given by
\begin{equation}\label{Z2invariant}
(-1)^{\Delta}=\prod_{\alpha=1,\cdots,4}\delta_{{\bf K}_{\alpha}}
\end{equation}

Eq. (\ref{Z2invariant}) defines the $Z_2$ topological invariant of magnetic systems with anti-unitary symmetry $A$. However, this definition is not convenient for the practical calculation in a realistic model since a global gauge should be chosen for the wave function $|\psi_{n,{\bf k}_\alpha}\rangle$. This difficulty can be overcome by considering the Wilson loop method \cite{yu2011,kingsmith1993}. This is because bulk topological invariant can be interpreted through a simple physical picture, partner switching of Wannier center flow \cite{fu2006}. Let's assume the system has four $A$-invariant momenta ${\bf K}_\alpha$ ($\alpha=1a,1b,2a,2b$), which fall in two parallel lines $\tilde{\bf K}_1$ and $\tilde{\bf K}_2$ (${\bf K}_{1a,1b}\in\tilde{\bf K}_1$ and ${\bf K}_{2a,2b}\in\tilde{\bf K}_2$). For simplicity, we can assume these two lines are along $k_z$ direction, which can be satisfied in all our cases by choosing appropriate coordinates. We can denote ${\bf K}_\alpha=(\vec{\kappa}_\alpha,k_z)$ where $\alpha=1,2$ and $\vec{\kappa}_\alpha=(k_{\alpha,x},k_{\alpha,y})$, so $\vec{\kappa}_\alpha$ is fixed for each line. Similar to the case of TR invariant TIs, we focus on doubly degenerate states or doublets. For each doublet, we can pick out one state $|\psi_{n,{\bf k}_\alpha}\rangle$ and define $|\psi^I_{n,{\bf k}_\alpha}\rangle=|\psi_{n,{\bf k}_\alpha}\rangle$ and $|\psi^{II}_{n,{\bf k}_\alpha}\rangle=A|\psi_{n,{\bf k}_\alpha}\rangle$, where the integer $n=1,2,\cdots$ is for all occupied doublet states. For the fixed $k_x$ and $k_y$, one can treat the system as a 1D system along the z direction and define a partial polarization for these doublet pairs as
\begin{equation}
	P^{i}(\vec{\kappa})=\sum_n\frac{1}{2\pi}\int_{-\pi}^{\pi}\langle\psi^i_{n,{\bf k}}|i\partial_{k_z}|\psi^i_{n,{\bf k}}\rangle d k_z\ \ (i=I,II).
\end{equation}
The partial polarization $P^{i}$ is directly related to the Wannier function center $\theta$ of the occupied states $|\psi^i_{n,{\bf k}_\alpha}\rangle$ by $\theta_i(\vec{\kappa})=2\pi P^{i}(\vec{\kappa})$. With this definition, the Wannier function centers are periodic with $2\pi$. The Wannier function centers, as well as the partial polarizations, can be derived based on the Wilson loop technique. Allowing $k_z$ to take discrete values, we define a $U(2N)$ Wilson loop for the fixed $k_x$ and $k_y$ \cite{yu2011},
\begin{equation}
	D(\vec{\kappa})=S_{0,1}S_{1,2}S_{2,3}...S_{N_{z}-2,N_{z}-1}S_{N_{z}-1,0}
 \end{equation}
Here, $2N\times2N$ overlap matrices $S_{m,n}$ are defined using the periodic parts of Bloch wave-functions:
 \begin{eqnarray}
	 S_{i,i+1}^{m,n}(\vec{\kappa})& = & \langle m, k_{z,i},\vec{\kappa}|n, k_{z,i+1},\vec{\kappa}\rangle \nonumber\\
 k_{z,i} & = & \frac{2\pi i}{N_{z} a}
 \end{eqnarray}
 where $i\in 0,1,...,N_{z}-1$. Then the phases of the eigenvalues of this $U(2N)$ matrix $D(\vec{\kappa})$ give us the Wannier function centers $\theta(\vec{\kappa})$ of the occupied bands.

Similar to the case of TR invariant TIs, the difference between two partial polarizations $n_{\vec{\kappa}_{\alpha}}=P^I(\vec{\kappa}_\alpha)-P^{II}(\vec{\kappa}_\alpha)$ can only be an integer due to $A$ symmetry. The $Z_2$ topological invariant is related to $n_{\vec{\kappa}_{\alpha}}$ by
\begin{equation}\label{Z2invariant2}
	\Delta=n_{\vec{\kappa}_{1}}-n_{\vec{\kappa}_{2}}\ \ mod\ 2.
\end{equation}
Therefore, by evaluating the Wannier funtion centers with the Wilson loop method, one can obtain the $Z_2$ topological number in a magnetic system with an anti-unitary $A$ symmetry. Below, we will first apply our general theory to two concrete models of TMCIs and then extend the discussion to general magnetic crystal structures.

\section{Tight-binding models}

\subsection{$C_4\Theta$ model}
\begin{figure}
   \begin{center}
      \includegraphics[width=3.5in,angle=0]{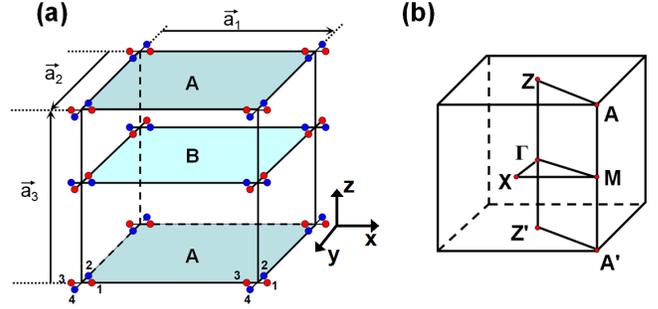}
    \end{center}
    \caption{ (a) Crystal structure of $C_{4}\Theta$ invariant tight-binding model. Atoms in blue and red have opposite z-directional magnetization. (b) The bulk Brillouin zone for our tight-binding model.}
    \label{fig1}
\end{figure}
Our first model for TMCIs possesses the combined symmetry of TR $\hat{\Theta}$ and four-fold rotation symmetry $\hat{C}_4$. Consider a layered magnetic structure along the z direction with a square lattice in the x-y plane. At one lattice site, there are four atoms with z-directional magnetic moments and the magnetic moments between two neighboring atoms are opposite, as shown in Fig. \ref{fig1}(a). In one layer, there are two sub-layers labeled by A and B. In-plane lattice vectors are given by $\vec{a}_1=(a,0,0)$ and $\vec{a}_2=(0,a,0)$, and the layers are stacked along $\vec{a}_3=(0,0,c)$. The distance between two neighboring sub-layers is denoted as $d$.

First, let us analyze symmetry properties of our system and show that a 2D irreducible co-representation could exist in our system. For the lattice structure shown in Fig. \ref{fig1}, due to the opposite alignment of magnetization at the red and blue sites, both TR symmetry $\hat{\Theta}$ and four-fold rotation symmetry $\hat{C}_{4}$ are broken. However, the combined symmetry $\hat{\Pi}=\hat{C}_{4}\hat{\Theta}$ exists for our system. It can be easily checked that the symmetries in our system are : $\hat{E}$, $\hat{C}_{4}\hat{\Theta}$, $\hat{C}_{4}^{3}\hat{\Theta}$, $\hat{C}_{2}$, $\hat{m}_{x}\hat{\Theta}$, $\hat{m}_{y}\hat{\Theta}$, $\hat{m}_{110}$  and $\hat{m}_{1-10}$. Here $\hat{m}_{x}$, $\hat{m}_{y}$, $\hat{m}_{110}$ and $\hat{m}_{1-10}$ are mirror reflections with the following mirror planes: (100), (010), (110) and (1-10). Therefore, the point symmetries of our system belong to the magnetic group $C_{4v}(C_{2v})$, where $C_{4v}$ is the magnetic point symmetry group and $C_{2v}$ is its unitary part. In fact, mirror symmetries are not the essential symmetry in our system, which will be shown later by including terms that break mirror symmetries. Thus, for the analysis below, we will ignore mirror symmetries and only consider $C_{4}(C_{2})$ group for simplicity.
The unitary part $C_2$ contains two 1D irreducible representations, A and B, the characters of which are given by $\chi_A(\hat{E})=\chi_B(\hat{E})=1$ and $\chi_A(\hat{C}_2)=-\chi_B(\hat{C}_2)=1$.
To see the reducibility of the co-representation of $C_{4}(C_{2})$, we could calculate $\sum_{{\bf B}\in {\bf G}\Theta}\chi(B^2)$ based on the Herring rule (\ref{eq:MSG_Herring1}). In $C_{4}(C_{2})$, ${\bf B}\in {\bf G}\Theta$ can be either $\hat{C_{4}}\hat{\Theta}$ or $\hat{C_{4}^{3}}\hat{\Theta}$. In both cases, ${\bf B}^2=\hat{C}_{2}$. Then it is easy to show that
\begin{equation}
\left.
\begin{aligned}
\sum_{{\bf B}\in {\bf G}\Theta}\chi_A(B^2)&=&2 \ \text{ case a}, \nonumber \\
\sum_{{\bf B}\in {\bf G}\Theta}\chi_B(B^2)&=&-2 \ \text{ case b}.   \nonumber
\end{aligned}
\right.
\label{C4(C2)class}
\end{equation}
So for the 1D representation B ($\Delta_B(\hat{C}_{2})=-1$), $C_{4}(C_{2})$ possesses a 2D irreducible co-representation which belongs to case b. Especially, the 2D irreducible co-representation for $\hat{C}_{4}\hat{\Theta}$ is
\begin{equation}
 \begin{aligned}
D_B(\hat{C}_{4}\hat{\Theta}) & = & \begin{bmatrix}
     0 & -1 \\
      1 & 0 \\
\end{bmatrix}
\end{aligned}.
\label{C4T_corep}
\end{equation}
This implies that it is possible to achieve double degeneracies in our $\hat{C}_{4}\hat{\Theta}$ model.

From the general co-representation theoretical analysis, we have shown the $\hat{C}_{4}\hat{\Theta}$ symmetry in our system could support doubly degenerate states. Next we will show the tight-binding Hamiltonian for our system. Let us consider a spinless model with $p_x$ and $p_y$ orbitals for each atom. The Hamiltonian is given by
\begin{widetext}
\begin{eqnarray}\label{C4Tmodel}
H&=&H_{A}+H_{B}+H_{AB} \nonumber \\
H_{\eta}&=&\sum_{i\neq j\in\{1,2,3,4\},\vec{m},\alpha,\beta,n}
t^{\alpha,\beta}_{\vec{m}^{\eta}_i,\vec{m}^{\eta}_j}c^{\eta\dagger}_{\alpha}(\vec{m}_i,n) c^{\eta}_{\beta}(\vec{m}_j,n)
+ \sum_{i\in\{1,2\},\vec{m},\alpha,\beta,n}[t^{\alpha,\beta}_{\vec{m}^{\eta}_i,\vec{m}^{\eta}_{i+2}+\vec{a}_i}
c^{\eta\dagger}_{\alpha}(\vec{m}_i,n)c^{\eta}_{\beta}(\vec{m}_{i+2}+\vec{a}_i,n) \nonumber \\
&+&t^{\alpha,\beta}_{\vec{m}^{\eta}_{i+2},\vec{m}^{\eta}_{i}-\vec{a}_i}
c^{\eta\dagger}_{\alpha}(\vec{m}_{i+2},n)c^{\eta}_{\beta}(\vec{m}_{i}-\vec{a}_i,n)]
+\sum_{i\in\{1,2,3,4\},\vec{m},n}\eta M_1[-ic^{\eta\dagger}_{p_x}(\vec{m}_i,n)c^{\eta}_{p_y}(\vec{m}_i,n)
+ic^{\eta\dagger}_{p_y}(\vec{m}_i,n)c^{\eta}_{p_x}(\vec{m}_i,n)] \nonumber \\
H_{AB}&=&\sum_{i,j\in\{1,2,3,4\},\vec{m},\alpha,\beta,n}t^{\alpha,\beta}_{\vec{m}^A_i,\vec{m}^B_j}
c^{A\dagger}_{\alpha}(\vec{m}_i,n)c^{B}_{\beta}(\vec{m}_j,n)
+\sum_{i\in\{1,2\},\vec{m},\alpha,\beta,n}[t^{\alpha,\beta}_{\vec{m}^{A}_i,\vec{m}^{B}_{i+2}+\vec{a}_i}
c^{A\dagger}_{\alpha}(\vec{m}_i,n)c^{B}_{\beta}(\vec{m}_{i+2}+\vec{a}_i,n) \nonumber \\
&+&t^{\alpha,\beta}_{\vec{m}^{A}_{i+2},\vec{m}^{B}_{i}-\vec{a}_i}
c^{A\dagger}_{\alpha}(\vec{m}_{i+2},n)c^{B}_{\beta}(\vec{m}_{i}-\vec{a}_i,n)]
+\sum_{i,j\in\{1,2,3,4\},\vec{m},\alpha,\beta,n}t^{\alpha,\beta,int}_{\vec{m}^A_i,\vec{m}^B_j}
c^{A\dagger}_{\alpha}(\vec{m}_i,n)c^{B}_{\beta}(\vec{m}_j,n+1)
\end{eqnarray}
\end{widetext}
where $\eta=A,B$ for sub-layer A and B. $M_1$ is the coupling between magnetic moments and p orbitals.  $(\vec{m},n)=(m_x,m_y,n)$ denotes lattice sites in the layer $n$ and $\alpha,\beta=p_x,p_y$ denote atomic orbitals. There are four atoms in one lattice site $\vec{m}$, as indicated by the lower indices $i$ and $j$. These four atoms locate at $\vec{m}_1=(m_x+b,m_y)$, $\vec{m}_2=(m_x,m_y+b)$, $\vec{m}_3=(m_x-b,m_y)$, $\vec{m}_4=(m_x,m_y-b)$. $t^{\alpha,\beta}_{\vec{m}^{\eta}_i,\vec{m}^{\eta}_j}$ is the on-site hopping amplitude between four atoms at one lattice site (For simplicity, we denote it as $t_1$). $t^{\alpha,\beta}_{\vec{m}^{\eta}_i,\vec{m}^{\eta}_{i+2}+\vec{a}_i}$ and $t^{\alpha,\beta}_{\vec{m}^{\eta}_{i+2},\vec{m}^{\eta}_{i}-\vec{a}_i}$ are the nearest-neighboring hopping coefficients in sub-layer $\eta$ (denoted as $t_3$). $t^{\alpha,\beta}_{\vec{m}^A_i,\vec{m}^B_j}$ is the nearest-neighboring hopping between the sub-layers A and B in the same layer $n$ (denoted as $t_2$). $t^{\alpha,\beta}_{\vec{m}^{A}_i,\vec{m}^{B}_{i+2}+\vec{a}_i}$ and $t^{\alpha,\beta}_{\vec{m}^{A}_{i+2},\vec{m}^{B}_{i}-\vec{a}_i}$ are the next-nearest-neighboring hopping coefficients between the sub-layers A and B in the same layer $n$ (denoted as $t_4$). $t^{\alpha,\beta,int}_{\vec{m}^A_i,\vec{m}^B_j}$ is the nearest-neighboring inter-layer hopping from the sub-layer A in layer $n$ to sub-layer B in layer $n+1$ (denoted as $t_5$).

The Bloch Hamiltonian $H({\bf k})$ is given in Appendix B, which is a $16\times 16$ matrix written under the basis $|i^{\eta}\alpha,{\bf k}\rangle=\frac{1}{\sqrt{N}}\sum_{n}e^{i\vec{k}\cdot {\vec{r}}^n_{\eta,i}}\phi_{\alpha}(\vec{r}-{\vec{r}}^n_{\eta,i})$. Here $N$ is the normalization factor. The position of each atom is shown by $\vec{r}^n_{\eta,i}=\vec{t}_n+\vec{r}_{\eta,i}$ with $\vec{t}_n$ denoting the lattice vector, $\eta=A,B$ denoting the sub-layers and $i\in\{1,2,3,4\}$ denoting the index of each atom. $\phi_{\alpha}$ denotes the wave function of the basis with $\alpha=p_x,p_y$.

Next we would like to apply the above general symmetry analysis to the Hamiltonian $H({\bf k})$. To do this, we need first to construct symmetry operations in the basis $|i^{\eta}\alpha,{\bf k}\rangle$ for our model. The essential anti-unitary operator in our system is the combined symmetry $\hat{\Pi}=\hat{C}_4\hat{\Theta}$, which can serve as the generator of the symmetry group.
Since our model is spinless, $\hat{\Theta}$ operation is just complex conjugate. The behavior of basis wave-functions under $\hat{\Pi}$ operation is given by
\begin{equation}
\hat{\Pi}|i^{\eta}\alpha,{\bf k}\rangle=
\sum_{j^{\eta}\beta}({\cal C}_4)^{\eta}_{i\alpha,j\beta}|j^{\eta}\beta,\hat{\Pi}{\bf k}\rangle
\end{equation}
where ${\cal C}_4^{\eta}$ is a block matrix defined for the sub-layer $\eta=A,B$, given by
\begin{equation}
	{\cal C}_4  =  \begin{bmatrix}
		{\cal C}_4^{A} & 0 \\
		0 & {\cal C}_4^{B} \\
\end{bmatrix}
\nonumber \\
\end{equation}
where
\begin{equation}
{\cal C}_4^{\eta} =  \begin{bmatrix}
     0 & U & 0 & 0 \\
      0 & 0 & U & 0 \\
      0 & 0 & 0 & U \\
      U & 0 & 0 & 0 \\
\end{bmatrix},
U  =  \begin{bmatrix}
     0 & -1 \\
      1 & 0 \\
\end{bmatrix}.
\nonumber 
\end{equation}
Here, the $2\times2$ matrix $U$ is defined on the basis $p_x$ and $p_y$. ${\cal C}_4^{\eta}$ gives the rotation matrix of four atoms at one site in the sub-layer $\eta$. 

By acting the above symmetry operations on our Bloch Hamiltonian $H({\bf k})$, we find that the Hamitlonian satisfies the combined symmetry $\hat{\Pi}$,
\begin{eqnarray}\label{Pisymmetry}
\hat{\Pi} H(k_x,k_y,k_z) \hat{\Pi}^{-1}&=& H(-k_y,k_x,-k_z).
\end{eqnarray}
According to Eq. (\ref{Pisymmetry}), it is easy to see that the Hamiltonian commutes with the operator $\hat{\Pi}$ at four $\hat{\Pi}$ invariant momenta $\Gamma$, Z, A, M, shown in Fig. \ref{fig1} (b). Therefore, all the eigenstates at these $\hat{\Pi}$-invariant momenta can be classified by the co-representation of $C_4(C_2)$ group.

It is convenient to first identify which co-representation the basis $|i^{\eta}\alpha,{\bf K}\rangle$ belongs to, for ${\bf K}=\Gamma,Z,A,M$. This is because the coupling matrix elements of the Hamiltonian must be zero for two basis that belong to different co-representations. From the general theory above, the co-representations of $C_{4}(C_{2})$ are classified by the eigenvalues of $\hat{C}_{2}$ (Eq. \ref{C4(C2)class}). Since $[H({\bf K}),\hat{C}_{2}]=0$, $H({\bf K})$ and $\hat{C}_{2}$ have common eigen-states. Thus, every energy eigen-state is also labeled by an eigenvalue of $\hat{C}_{2}$. If the $\hat{C}_{2}$ eigenvalue of a state is $-1$, this state belongs to a doublet pair at $\hat{\Pi}$ invariant momenta. We would expect an eigen-state with a $\hat{C}_{2}$ eigenvalue of $-1(+1)$ could be written as a linear combination of the corresponding basis wave functions. We perform a unitary transformation of basis wave-functions and find two sets of linear combinations of basis functions:
\begin{equation}
\left\{
\begin{aligned}
|\Phi^{\eta}_{s1,{\bf k}}\rangle_{1}=|1^{\eta}p_x,{\bf k}\rangle+|2^{\eta}p_y,{\bf k}\rangle
-|3^{\eta}p_x,{\bf k}\rangle-|4^{\eta}p_y,{\bf k}\rangle \nonumber \\
|\Phi^{\eta}_{s2,{\bf k}}\rangle_{1}=|1^{\eta}p_x,{\bf k}\rangle-|2^{\eta}p_y,{\bf k}\rangle
-|3^{\eta}p_x,{\bf k}\rangle+|4^{\eta}p_y,{\bf k}\rangle \nonumber \\
|\Phi^{\eta}_{d1,{\bf k}}\rangle_{1}=|1^{\eta}p_x,{\bf k}\rangle+|2^{\eta}p_y,{\bf k}\rangle
+|3^{\eta}p_x,{\bf k}\rangle+|4^{\eta}p_y,{\bf k}\rangle  \nonumber \\
|\Phi^{\eta}_{d2,{\bf k}}\rangle_{1}=|1^{\eta}p_x,{\bf k}\rangle-|2^{\eta}p_y,{\bf k}\rangle
+|3^{\eta}p_x,{\bf k}\rangle-|4^{\eta}p_y,{\bf k}\rangle \nonumber
\end{aligned}
\right.
\end{equation}
and
\begin{equation}
\left\{
\begin{aligned}
|\Phi^{\eta}_{s1,{\bf k}}\rangle_{2}=|1^{\eta}p_y,{\bf k}\rangle+|2^{\eta}p_x,{\bf k}\rangle
-|3^{\eta}p_y,{\bf k}\rangle-|4^{\eta}p_x,{\bf k}\rangle \nonumber \\
|\Phi^{\eta}_{s2,{\bf k}}\rangle_{2}=|1^{\eta}p_y,{\bf k}\rangle-|2^{\eta}p_x,{\bf k}\rangle
-|3^{\eta}p_y,{\bf k}\rangle+|4^{\eta}p_x,{\bf k}\rangle \nonumber \\
|\Phi^{\eta}_{d1,{\bf k}}\rangle_{2}=|1^{\eta}p_y,{\bf k}\rangle+|2^{\eta}p_x,{\bf k}\rangle
+|3^{\eta}p_y,{\bf k}\rangle+|4^{\eta}p_x,{\bf k}\rangle  \nonumber \\
|\Phi^{\eta}_{d2,{\bf k}}\rangle_{2}=|1^{\eta}p_y,{\bf k}\rangle-|2^{\eta}p_x,{\bf k}\rangle
+|3^{\eta}p_y,{\bf k}\rangle-|4^{\eta}p_x,{\bf k}\rangle  \nonumber
\end{aligned}
\right.
\end{equation}
where the indices s and d represent singlets and doublets at $\hat{\Pi}$-invariant momenta, as we can see from their $\hat{C}_{2}$ eigenvalues,
\begin{equation}
\left\{
\begin{aligned}
\hat{C}_{2}|\Phi^{\eta}_{s1,{\bf K}}\rangle_{\delta}&=&|\Phi^{\eta}_{s1,{\hat{C}_{2}\bf K}}\rangle_{\delta}  \nonumber \\
\hat{C}_{2}|\Phi^{\eta}_{s2,{\bf K}}\rangle_{\delta}&=&|\Phi^{\eta}_{s2,{\hat{C}_{2}\bf K}}\rangle_{\delta}  \nonumber \\
\hat{C}_{2}|\Phi^{\eta}_{d1,{\bf K}}\rangle_{\delta}&=&-|\Phi^{\eta}_{d1,{\hat{C}_{2}\bf K}}\rangle_{\delta} \nonumber \\
\hat{C}_{2}|\Phi^{\eta}_{d2,{\bf K}}\rangle_{\delta}&=&-|\Phi^{\eta}_{d2,{\hat{C}_{2}\bf K}}\rangle_{\delta}
\end{aligned}
\right.
\end{equation}
where ${\bf K}=\Gamma,Z,A,M$. The indices $\delta=1,2$ are related to the combined symmetry operator $\hat{\Pi}$ by
\begin{equation}
\left\{
\begin{aligned}
\hat{\Pi}|\Phi^{\eta}_{d1,{\bf k}}\rangle_{\delta}&=&(-1)^{\delta}|\Phi^{\eta}_{d2,{\hat{\Pi}\bf k}}\rangle_{\delta} \nonumber \\
\hat{\Pi}|\Phi^{\eta}_{d2,{\bf k}}\rangle_{\delta}&=&-(-1)^{\delta}|\Phi^{\eta}_{d1,{\hat{\Pi}\bf k}}\rangle_{\delta}
\end{aligned}
\right.
\end{equation}
At $\hat{\Pi}$ invariant momenta, the relations above give us exactly the 2D irreducible co-representation of $\hat{C}_{4}\hat{\Theta}$ (Eq. (\ref{C4T_corep})).

\begin{figure}
   \begin{center}
      \includegraphics[width=3.5in,angle=0]{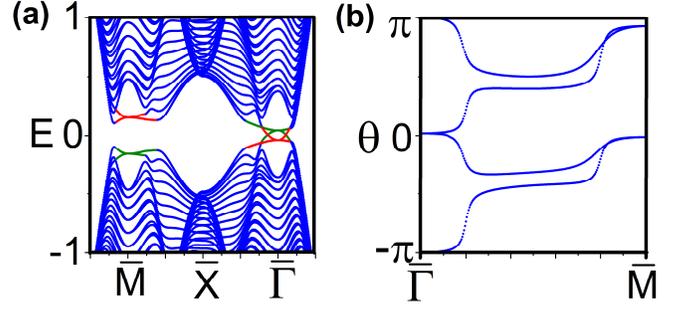}
    \end{center}
    \caption{ (a) Energy dispersion of a slab geometry configuration with the parameter $t_{p1} = -0.4, t_{p2} = -0.3, t_{s1} = 0.6,t_{s2} = 0.3,t_{p3} = -0.1,t_{p4} = 1,t_{s3} = 0.25,t_{s4} = 3.5,M_1 = 4.5,t_{p5} = 3.3,t_{s5} = 3.5$ (Please see Appendix B for definitions of coefficients here). Red and green lines represent surface states on top and bottom surfaces. The corresponding evolution of Wannier function centers is shown in (b), where topological invariant $\Delta=1$.}
    \label{sfwc}
\end{figure}

The above symmetry analysis has indicated that topological phases protected by $\hat{\Pi}$ symmetry could exist in our model Hamiltonian. To confirm this phase, we perform numerical calculations of topological surface states, as well as topological invariants. We carry out an energy dispersion calculation for our model Eq.(\ref{C4Tmodel}) in a slab geometry along the z direction, as shown in Fig. \ref{sfwc}(a). By choosing appropriate parameters, we find that surface states appear at the $\Gamma$ point of the surface BZ in the bulk energy gap, as shown in Fig. \ref{sfwc}(a), where red and green lines are the dispersions of surface states at top and bottom surfaces, respectively. For each surface, there are two branches of surface states, which connect conduction bands to valence bands and touch at the $\Gamma$ point quadratically. The analysis of the touching point shows that the surface wave functions belong to the 2D irreducible co-representation of $C_{4}(C_{2})$ group. Therefore, the gapless surface states are indeed protected by $\hat{\Pi}$ symmetry.

In addition, we calculate bulk topological invariants of our model to confirm the topological nature of surface states. Since there are four $\hat{\Pi}$ invariant momenta $\Gamma,Z,A,M$, according to the discussion in Sec. IIC, one $Z_2$ topological invariant can be defined by Eq. (\ref{Z2invariant}) or by Eq. (\ref{Z2invariant2}) based on the Wannier function centers.
In our system, the Wannier function centers can be defined along the $k_z$ direction for any fixed $k_x$ and $k_y$. We can track the evolution of the Wannier function centers from $k_{x,y}=0$ (denoted as $\bar{\Gamma}$) to $k_{x,y}=\pi$ (denoted as $\bar{M}$), as shown in Fig. \ref{sfwc}(b). Four Wannier function centers in Fig. \ref{sfwc}(b) result from two doublet states in our model. As expected, all Wannier function centers are degenerate at both $\bar{\Gamma}$ and $\bar{M}$ due to $\hat{\Pi}$ symmetry. More importantly, we find that the degenerate pairs of the Wannier function centers switch their partners once between $\bar{\Gamma}$ and $\bar{M}$, indicating the $Z_2$ number $\Delta=1$ for the Eq. (\ref{Z2invariant2}). This result agrees with the existence of gapless surface states in Fig. \ref{sfwc}(a).

\begin{figure}
   \begin{center}
      \includegraphics[width=3.5in,angle=0]{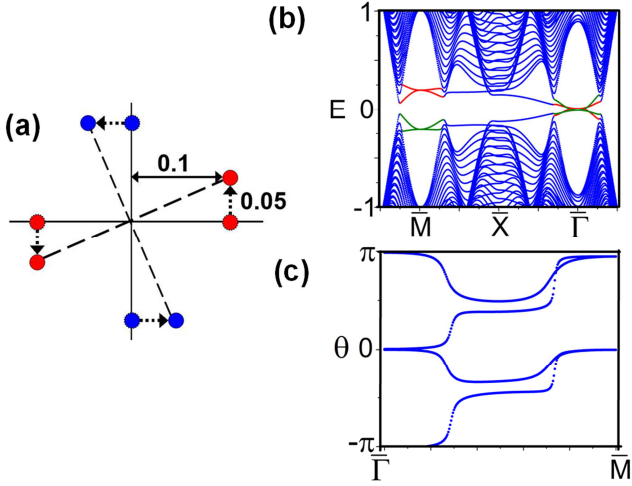}
    \end{center}
    \caption{After performing a rotation operation to break mirror symmetry (as is shown in (a)), surface state (b) is found in the following parameter: $t_{p1} = -0.4, t_{p2} = -0.3, t_{s1} = 0.6,t_{s2} = 0.3,t_{p3} = -0.1,t_{p4} = 1,t_{s3} = 0.25,t_{s4} = 5,M_1 = 4.5,t_{p5} = 3.3,t_{s5} = 3.5$. The Wannier center flow plotted in (c) shows a non-trivial topological invariant.}
    \label{bkmir}
\end{figure}

Mirror symmetry is claimed not to be essential to protect double degeneracies and we will discuss it below. Notice that mirror-invariant plane in our model is (1-10) plane, where four $\hat{C_4}\hat{\Theta}$ invariant momenta ($\Gamma$, $M$, $A$, $Z$) locate. This makes it possible to define a mirror Chern number in our system \cite{hsieh2012}. To rule out this possibility, we perform a rotation operation to the four atoms at one lattice site, as shown in Fig. \ref{bkmir}(a). This new configuration of atoms breaks the following mirror symmetries: $\hat{m}_{x}\hat{\Theta}$, $\hat{m}_{y}\hat{\Theta}$, $\hat{m}_{110}$  and $\hat{m}_{1-10}$. The only remaining symmetries form exactly the $C_{4}(C_{2})$ magnetic group. We calculate the energy dispersion in a slab geometry along the z direction (Fig.\ref{bkmir} (b)), and find that gapless surface states still show up. These surface states are degenerate at exactly $\bar{\Gamma}$, which implies they are only protected by $\hat{C_4}\hat{\Theta}$ symmetry. In Fig. \ref{bkmir}(c), we plot the Wannier function center evolution in this new mirror-breaking model. As we expect, a non-trivial topological invariant $\Delta=1$ appears.

Finally, we would like to comment on the connection between the Fu's model with both $\hat{C}_{4}$ and $\Theta$ symmetry in Ref. \cite{fu2011a} and our $\hat{C}_{4}\hat{\Theta}$ model. In Fu's model, $\hat{C}_{4}$ rotation symmetry and time reversal symmetry are preserved separately, and double degeneracies of gapless surface states also appear at $\Gamma$ \cite{fu2011a}. Based on the co-representation theory and the Herring rule, we find that the essential symmetry in Fu's model is also the combined symmetry $\hat{C}_{4}\hat{\Theta}$. Therefore, our $\hat{C}_{4}\hat{\Theta}$ model is a generalization of Fu's model to magnetic structures. As pointed out by Fu\cite{fu2011a}, if singlet states appear near the Fermi energy, they will weaken the stability of topological surface states consisting of doublet pairs. Similar situations will occur in our model.


\subsection{$\hat{{\bf \tau}}\hat{\Theta}$ model}

\begin{figure}
   \begin{center}
      \includegraphics[width=3.5in,angle=0]{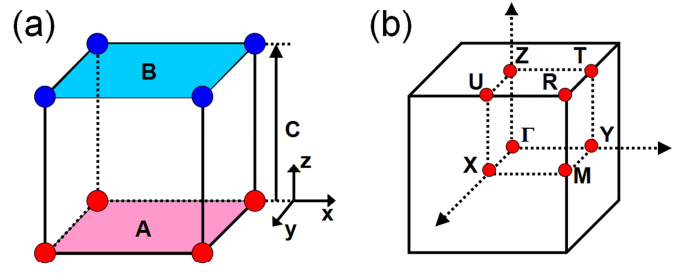}
    \end{center}
    \caption{(a) Layered lattice structure of $\hat{{\bf \tau}}\hat{\Theta}$ model. Atoms with different colors have opposite z-directional magnetic moments. Layer A and Layer B are related by a translation of a vector {\bf c}. (b) BZ of $\hat{{\bf \tau}}\hat{\Theta}$ model and eight $\hat{{\bf \tau}}\hat{\Theta}$ invariant momenta.}
    \label{lattice_ThetaT}
\end{figure}

The above $\hat{C_4}\hat{\Theta}$ protected topological phases could exist in magnetic structures with type-IIIa magnetic space symmetry groups, in which the combination of point symmetry and TR symmetry plays an essential role. Similar topological phases can also be found in magnetic structures with type-IIIb magnetic space symmetry groups. As shown in Ref. \cite{mong2010,fang2013}, $Z_2$ topological phases also exist in anti-ferromagnetic models with surface states protected by the combination of the TR operation $\hat{\Theta}$ and the translation operation $\hat{{\bf \tau}}$. However, these models contain spins and are related to the magnetic double groups. In contrast, we consider an anti-ferromagnetic model with spinless fermions in Ref. \cite{liu2013c}, which will be briefly reviewed below. In this model, 2D ferromagnetic square lattices are layered along z direction with opposite z-directional magnetic moments in neighboring layers, as shown in Fig. \ref{lattice_ThetaT}(a). So the system has an anti-ferromagnetic structure with two atoms (A and B in Fig. \ref{lattice_ThetaT}(a)) in one unit cell. Three different orbitals, $|s\rangle$, $|p_x\rangle$ and $|p_y\rangle$, are considered, and the tight-binding Hamiltonian is given in Ref. \cite{liu2013c}.

Due to the arrangement of magnetic moments, the symmetry group of this model can be generated by the combined symmetry of $\hat{{\bf \tau}}\hat{\Theta}$, where $\hat{{\bf \tau}}$ is a translation operator by a vector ${\bf c}=(0,0,c)$ in z direction, denoted in Fig. \ref{lattice_ThetaT}(a). The corresponding unitary part of ths symmetry group is nothing but the translation group generated by the operator $\hat{\bf t}=(\hat{{\bf \tau}}\hat{\Theta})^2=(0,0,2c)$. The 1D representation of a translation symmetry group is labelled by a wave vector (or a momentum) $k_z=\frac{m\pi}{Nc}$ ($m$ and $N$ are integers), as shown in Table \ref{tb:character_ThetaT}. Here we have used a periodical boundary condition $\hat{\bf t}^N=\hat{E}$. Notice that the symmetry operation $\hat{{\bf \tau}}\hat{\Theta}$ only exists for two momenta, $k_z=0$ and $\frac{\pi}{2c}$.
%
\begin{table}
\caption{Character table of translation group T}
\begin{tabular}{c c c c c c}
\hline\hline
$k_{z}$ & $\hat{E}$ & $\hat{\bf t}$ & $\hat{\bf t}^2$ & ... & $\hat{\bf t}^{N-1}$ \\ [1ex]
\hline
$0$ & 1 & 1 & 1 & ... & 1 \\
$\frac{\pi}{Nc}$ & 1 & $e^{i\frac{2\pi}{N}}$ & $e^{i\frac{4\pi}{N}}$ & ... & $e^{i\frac{2(N-1)\pi}{N}}$ \\
$...$ & ... & ... & ... & ... & ... \\
$\frac{\pi}{2c}$ & 1 & -1 & 1 & ... & $(-1)^{N-1}$ \\
$...$ & ... & ... & ... & ... & ... \\
$\frac{(N-1)\pi}{Nc}$ & 1 & $e^{i\frac{2(N-1)\pi}{N}}$ & $e^{i\frac{4(N-1)\pi}{N}}$ & ... & $e^{i\frac{2(N-1)^2\pi}{N}}$\\[1ex]
\hline
\end{tabular}
\label{tb:character_ThetaT}
\end{table}
Based on the Herring rule (Eq. \ref{eq:MSG_Herring1}), one finds that
\begin{eqnarray}
\sum_{{\bf B}\in {\bf G}\Theta}\chi(B^2)&=&\sum_{m=0}^{N-1}\chi[(\hat{\bf t}^{m})^2] \nonumber \\
&=&\left\{
\begin{aligned}
N\  \text{   $k_z=0$, case a}  \\
0\  \text{    $k_z=\frac{\pi}{2c}$, case c}
\end{aligned}
\right.
\label{ThetaT_class}
\end{eqnarray}

So when $k_z=\frac{\pi}{2c}$, magnetic translation group is possible to have 2D irreducible co-representations (double degeneracies) at $\hat{{\bf t}}\hat{\Theta}$ invariant momenta $Z$, $U$, $R$, $T$ in Fig. \ref{lattice_ThetaT} (b). When $k_z=0$, at the other four $\hat{{\bf t}}\hat{\Theta}$ invariant momenta, $\Gamma$, $X$, $M$, $Y$ in Fig. \ref{lattice_ThetaT} (b), only singlet states exist. Thus, based on the co-representation theory, we arrive at the same conclusion as that of the previous analysis in Ref. \cite{liu2013c}.

\begin{figure}
   \begin{center}
      \includegraphics[width=3.5in,angle=0]{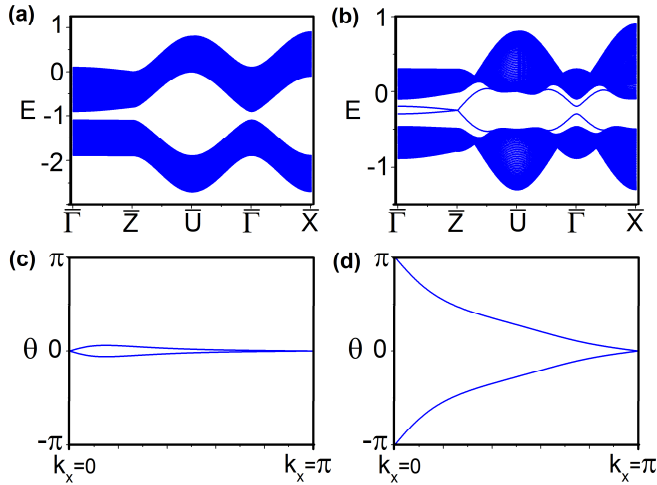}
    \end{center}
    \caption{Energy dispersion of $T\Theta$ model with a y-directional slab geometry configuration when (a) $M_1=3.1$ and (b) $M_1=4.5$. The corresponding evolution of Wannier function centers in a 3D bulk system is shown as a function of $k_x$ for (c) $M_1 = 3.1$, $\Delta=0$ and (d) $M_1 = 4.5$, $\Delta=1$.}
    \label{thetaT}
\end{figure}

In Fig. \ref{thetaT} (a) and (b), we calculate energy dispersions of $\hat{{\bf \tau}}\hat{\Theta}$ model in a slab geometry along y direction. In Fig. \ref{thetaT} (c) and (d), the corresponding Wannier center flows are plotted by calculating the evolution of 1D polarization $P(k_x,k_z=\frac{\pi}{2c})$ from $k_x=0$ to $k_x=\pi$. There is no gapless surface state in Fig. \ref{thetaT} (a) and a trivial topological invariant $\Delta=0$ in Fig. \ref{thetaT} (c), so the system belongs to a topologically trivial phase. In Fig. \ref{thetaT} (b) and (d), the system has a non-trivial topological invariant $\Delta=1$ and gapless surface states appear. Notice that surface states are gapless and degenerate at $\bar{Z}$ but not at $\bar{\Gamma}$. This is consistent with our above analysis based on the Herring rule.


\subsection{Other magnetic symmetry groups}
Next we will generalize our theory to other magnetic groups generated by $\hat{A}=\hat{\Theta} \hat{R}$, where $\hat{R}$ can be translation, mirror or rotation operations. The case of the translation symmetry has been well discussed for the $\hat{\bf \tau}\Theta$ model in the last section. For rotation operations, $\hat{R}=\hat{C}_n$ where $n$ can only be taken $n=2,3,4,6$ due to crystallographic restriction theorem. The Herring rule can be applied to both the single and double groups to extract the degeneracy induced by the anti-unitary operation $A$. Furthermore, we can identify the number of $\hat{A}$ invariant momenta in the BZ. Based on these results, we can identify which types of magnetic symmetry groups are able to host $Z_2$ topological phases. We will discuss different cases below separately.

\subsubsection{mirror symmetry}
We will first focus on the mirror symmetry. There are always two elements $E$ and $\hat{m}\hat{\Theta}$ for both the single and double group cases. One should note that for the double group case, $(\hat{m}\hat{\Theta})^2=\hat{m}^2\hat{\Theta}^2=\bar{E}\bar{E}=E$, where $\bar{E}$ for the rotation of $2\pi$ is different from the identity $E$ in a double group. Since the unitary part has only one element $E$, the system belongs to the case a and has no non-trivial topological phase.

\subsubsection{Rotation symmetry $n=2$}
Similar to the mirror symmetry, there are also only two elements $E$ and $\hat{C}_2\hat{\Theta}$ in this case, so no topological phase is possible.

\subsubsection{Rotation symmetry $n=3$}

\begin{figure}
   \begin{center}
      \includegraphics[width=3.5in,angle=0]{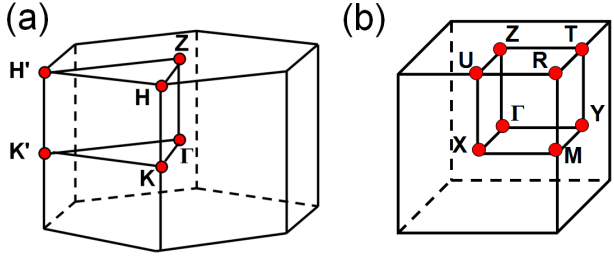}
    \end{center}
    \caption{ BZ and high symmetry points for $\hat{C}_n\hat{\Theta}$ lattice are drawn in: (a) n=3 and n=6 (b) n=4.}
    \label{BZ_Cn}
\end{figure}

There are six elements $E,\hat{C}_3,\hat{C}_3^2,\hat{\Theta},\hat{\Theta} \hat{C}_3,\hat{\Theta} \hat{C}_3^2$ for the $n=3$ and single group case. For the unitary part, three different 1D irreducible representations $\Gamma_i, i=1,2,3$ are listed in Table. \ref{tb:single_C3}, in which $\Gamma_1$ belongs to the case a while $\Gamma_{2,3}$ belong to the case c. However, systems with such symmetries have only two invariant momenta at $\Gamma$ and $Z$, as is shown in Fig \ref{BZ_Cn}(a). In order to define a $Z_2$ topological invariant, there should be at least four invariant momenta. Therefore, no $Z_2$ topological phase is possible in the $n=3$ and single group case. Similar argument can also be applied to the $n=3$ and double group case.

\begin{table}[ht]
\caption{Character table of $n=3$ in the single group case. Here $\omega=e^{\frac{\pi i}{3}}$.}
\begin{tabular}{c c c c c c}
\hline\hline
Rep. & $E$ & $C_{3}$ & $C_{3}^{2}$ & $\sum \chi(B^2)$ & case  \\ [1ex] \hline
$\Gamma_1$ &1 & 1 & 1 & 3 & a\\
$\Gamma_2$ &1 & $\omega$ & ${\omega}^2$ & 0 & c\\
$\Gamma_3$ &1 & ${\omega}^2$ & $\omega$ & 0 & c\\ [1ex]
\hline
\end{tabular}
\label{tb:single_C3}
\end{table}

\subsubsection{Rotation symmetry $n=4$}
The $\hat{C}_4\hat{\Theta}$ symmetry has been discussed in the $\hat{C}_4\hat{\Theta}$ model above for the single group case. Since there are four momenta invariant under $\hat{C}_4\hat{\Theta}$ operation in the BZ ( Fig. \ref{BZ_Cn} (b)), one $Z_2$ topological invariant can be defined for spinless fermions with the $\hat{C}_4\hat{\Theta}$ symmetry.

For the double group case, there are eight elements $E,\hat{\Theta} \hat{C}_4, \bar{E}\hat{C}_2,\hat{\Theta} \bar{E} \hat{C}_4^3, \bar{E}, \hat{\Theta}\bar{E}\hat{C}_4, \hat{C}_2, \hat{\Theta} \hat{C}_4^3$. Its unitary part is the double group of $\hat{C}_2$ symmetry, which possesses four 1D irreducible representations, denoted as $\Gamma_{1,2,3,4}$ in Table. \ref{tb:double_C4}. We find that $\Gamma_2$ belongs to case b and $\Gamma_{3,4}$ belongs to case c, so double degeneracy is also possible for $\hat{C}_4\hat{\Theta}$ symmetry in the double group case. Four $\hat{C}_4\hat{\Theta}$-invariant momenta( $\Gamma$, $Z$, $M$ and $R$) allow one $Z_2$ topological invariant.

\begin{table}[ht]
\caption{Character table of $n=4$, double group}
\begin{tabular}{c c c c c c c}
\hline\hline
Rep. & $E$ & $C_2$ & $\bar{E}$ & $C_2\bar{E}$ & $\sum \chi(B^2)$ & class  \\ [1ex] \hline
$\Gamma_1$ & 1 & 1 & 1 & 1 & 4 & a\\
$\Gamma_2$ &1 & -1 & 1 & -1 & -4 & b\\
$\Gamma_3$ &1 & i & -1 & -i & 0 & c\\
$\Gamma_4$ &1 & -i & -1 & i & 0 & c\\ [1ex]
\hline
\end{tabular}
\label{tb:double_C4}
\end{table}

\subsubsection{Rotation symmetry $n=6$}
The single group of $\hat{C}_6\hat{\Theta}$ symmetry contains six elements $E,\hat{\Theta} \hat{C}_6,\hat{C}_3,\hat{\Theta} \hat{C}_6^3, \hat{C}_3^2,\hat{\Theta} \hat{C}_6^5$. The unitary part in this case is the same as that of the $n=3$ case (single group). Making use of Table. \ref{tb:single_C3}, we have $\Gamma_1$ belonging to class a, and $\Gamma_{2,3}$ belonging to class c, so double degeneracy is possible for $\Gamma_2$ and $\Gamma_3$ representations. Different from $\hat{C}_3\hat{\Theta}$ symmetry, six momenta (Fig. \ref{BZ_Cn} (a)) are invariant under the $\hat{C}_6\hat{\Theta}$ operation. Then we could define one topological invariant (Eq. \ref{Z2invariant}) for each of the following planes that contains four $\hat{C}_6\hat{\Theta}$ invariant momenta:
\begin{eqnarray}
\text{Plane}\ \Gamma-Z-H'-K'&:&\ (-1)^{\Delta_{1}}=\delta_{\Gamma}\delta_{Z}\delta_{H'}\delta_{K'} \nonumber \\
\text{Plane}\ K'-H'-H-K&:&\ (-1)^{\Delta_{2}}=\delta_{K'}\delta_{H'}\delta_{H}\delta_{K} \nonumber \\
\text{Plane}\ \Gamma-Z-H-K&:&\ (-1)^{\Delta_{3}}=\delta_{\Gamma}\delta_{Z}\delta_{H}\delta_{K}
\end{eqnarray}

These three indices $\Delta_{i}(i=1,2,3)$, however, have to obey the following constraint :
\begin{eqnarray}\label{C6T_constrain}
(-1)^{\Delta_{1}+\Delta_{2}+\Delta_{3}}
&=& (\delta_{\Gamma}\delta_{Z}\delta_{H}\delta_{K}\delta_{H'}\delta_{K'})^2 \nonumber \\
&=& 1
\end{eqnarray}

\begin{figure}
   \begin{center}
      \includegraphics[width=3.5in,angle=0]{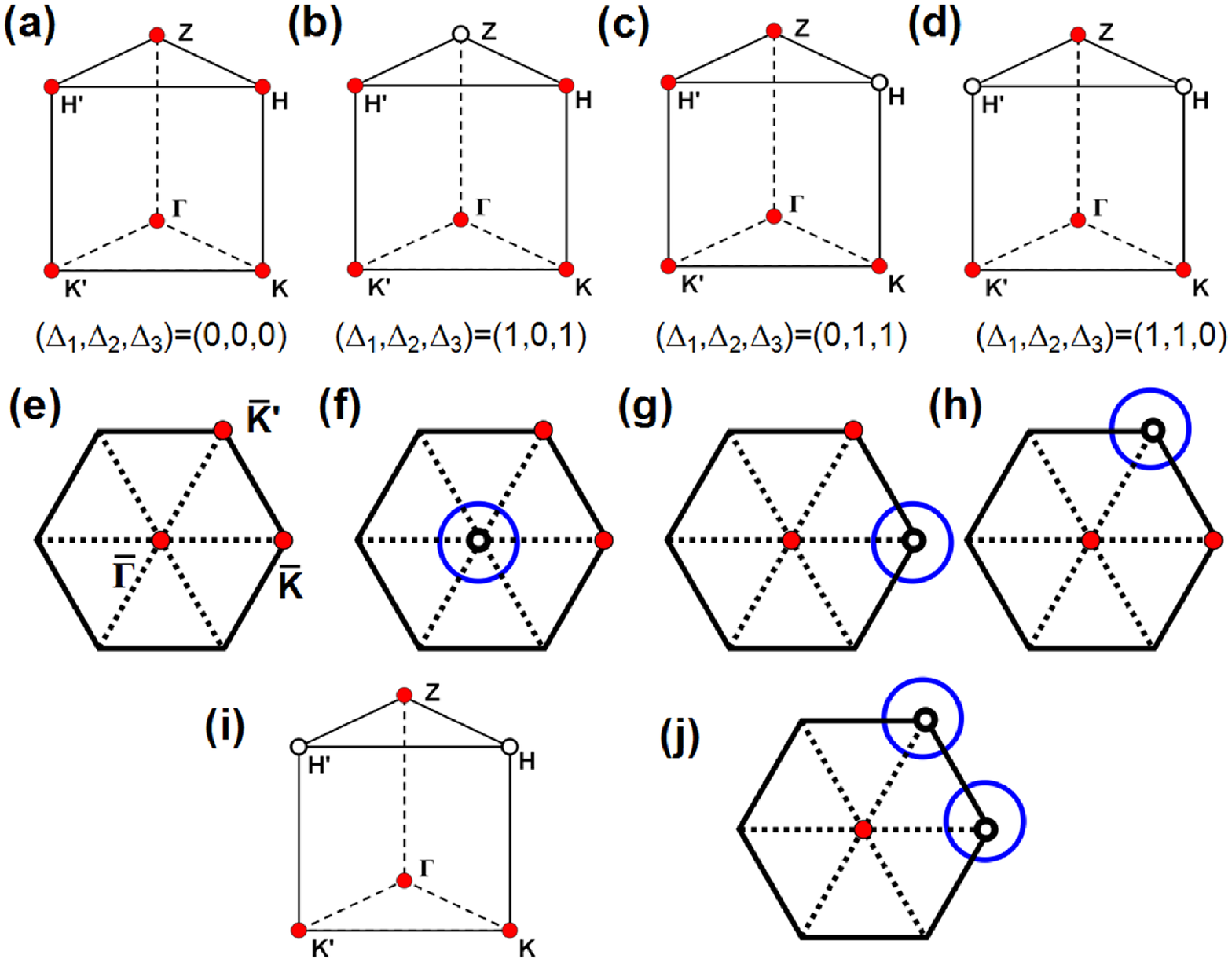}
    \end{center}
    \caption{Examples of four topologically inequivalent phases in a system with $\hat{C}_6\hat{\Theta}$ symmetry are shown in (a)-(d). (a) corresponds to a trivial phase while (b)-(d) give three distinct topologically non-trivial phases. The corresponding Fermi surfaces (blue circles) of surface states in (001) surface are shown in (e)-(h). In (i) and (j), we show another possible topological phase with an even number of gapless surface states, which is topologically equivalent to the case with an odd number of gapless surface states in (b) and (f). }
    \label{C6T_class}
\end{figure}

Consequently, three topologically non-trivial and inequivalent phases, as shown in Fig. \ref{C6T_class} (b)-(d), can exist in our $\hat{C}_6\hat{\Theta}$ model, besides the topologically trivial phase shown in Fig. \ref{C6T_class} (a). Here we use red spots at $\hat{C}_6\hat{\Theta}$ invariant momenta ${\bf K}_{\alpha}$ to represent $\delta_{{\bf K}_{\alpha}}=1$ while open circles to represent $\delta_{{\bf K}_{\alpha}}=-1$. To see their Fermi surface structures, we could define $\pi_{\bar{{\bf K}}_\alpha}=\delta_{{\bf K}_{\alpha1}}\delta_{{\bf K}_{\alpha2}}$, where ${\bf K}_{\alpha1}$ and ${\bf K}_{\alpha2}$ are $\hat{C}_6\hat{\Theta}$ invariant momenta that can be projected to $\bar{{\bf K}}_\alpha$ along (001) direction. Using the same coloring convention to show the value of $\pi_{\bar{{\bf K}}_\alpha}$, we arrive at the corresponding Fermi surfaces and surface BZ of (001) surface plotted in Fig. \ref{C6T_class} (e)-(h), where possible Fermi surfaces of surface band structures are shown in blue circles. Let us take the case where $(\Delta_1,\Delta_2,\Delta_3)=(1,0,1)$ as an example. In Fig. \ref{C6T_class} (b) and (f), $\pi_{\bar{\Gamma}}=\delta_{\Gamma}\delta_{Z}=-1$. This indicates one single Dirac-cone-like gapless surface structure at $\bar{\Gamma}$. As is shown in Fig. \ref{C6T_class} (f), Fermi energy will cross surface bands an even(odd) number of times along the path $\bar{K}-\bar{K'}$ ($\bar{\Gamma}-\bar{K}$ and $\bar{\Gamma}-\bar{K'}$). However, we also found another topologically equivalent phase with the same topological invariants and the same number of Fermi level crossing, but with two Dirac-cone-like gapless surface structures at $\bar{K}$ and $\bar{K'}$, shown in Fig. \ref{C6T_class} (i) and (j). It can be checked that by adding surface impurity bands, the case in Fig. \ref{C6T_class} (b) and (f) could be adiabatically connected to the case in Fig \ref{C6T_class} (i) and (j) without closing the band gap.

Constrained by Eq. \ref{C6T_constrain}, only two topological invariants are independent from each other. Different from the strong and weak indices for TR-invariant TIs, all three topologically non-trivial phases share the same stability. Therefore, without loss of generality, we can choose a $Z_2$ topological invariant pair $(\Delta_1,\Delta_2)$ to classify $Z_2\times Z_2$ topological phases in a $\hat{C}_6\hat{\Theta}$ model. When $(\Delta_1,\Delta_2)=(0,0)$, the system is topologically trivial. When $(\Delta_1,\Delta_2)=(0,1),(1,0),(1,1)$, the system is topologically non-trivial.

For the double group, there are also six elements: $\hat{C}_{6}\hat{\Theta}$, $\hat{C}_{6}^{2}\bar{E}$, $\hat{C}_{6}^{3}\bar{E}\hat{\Theta}$, $\hat{C}_{6}^{4}$, $\hat{C}_{6}^{5}\hat{\Theta}$, $E$ and the character table for the unitary part is shown in Table. \ref{tb:double_C6}. Similar to the single group case, double degeneracy exists for $\Gamma_{2,3}$ representations based on Table. \ref{tb:double_C6}. One $Z_2\times Z_2$ topological invariant pair $(\Delta_1,\Delta_2)$ could be defined among three $Z_2$ topological invariants $\Delta_{1,2,3}$ to classify the topological phases protected by $\hat{C}_6\hat{\Theta}$ double group symmetry.

\begin{table}[ht]
\caption{Character table of $n=6$, double group. Here $\omega=e^{\frac{2\pi i}{3}}$.}
\begin{tabular}{c c c c c c}
\hline\hline
Rep. & $E$ & $C_{6}^{4}$ & $\bar{E}C_{6}^{2}$ & $\sum \chi(B^2)$ & class  \\ [1ex] \hline
$\Gamma_1$ &1 & 1 & 1 & 3 & a\\
$\Gamma_2$ &1 & $\omega$ & ${\omega}^2$ & 0 & c\\
$\Gamma_3$ &1 & ${\omega}^2$ & $\omega$ & 0 & c\\ [1ex]
\hline
\end{tabular}
\label{tb:double_C6}
\end{table}

 Up to now, we have shown that besides Kramers' degeneracy due to TR symmetry $\hat{\Theta}$, systems with $\hat{C}_{4}\hat{\Theta}$ or $\hat{C}_{6}\hat{\Theta}$ symmetry operations are able to support topological phases for both spinless and spinful cases. This result can be applied to more complicated magnetic point groups and magnetic space groups when magnetic cyclic groups are their subgroups.



\section{Discussion and Conclusion}
Based on the Herring rule in co-representation theory of magnetic symmetry groups, we have developed a general theory for topological magnetic crystalline insulators. we have shown $Z_2$ topological phases can exist in a system with the $\hat{C}_4\hat{\Theta}$ or $\hat{\tau}\hat{\Theta}$ model while $Z_2\times Z_2$ topological phases can exist in a system with the $\hat{C}_6\hat{\Theta}$ symmetry. This result extends the previous discussions \cite{mong2010,fang2013,liu2013c} from magnetic translational group to magnetic point groups or magnetic space groups. One can easily extend our results to other magnetic symmetry groups with more than one generator.  To search for realistic magnetic materials, one should first identify crystal structures and magnetic structures of a system. For example, $\hat{C}_4$ symmetry exists in a large group of materials with cubic symmetry, such as face-centered cubic lattice, perovskite crystals, etc. In a cubic lattice, there are different antiferromagnetic structures, including A-type, C-type and G-type (e.g. see the Fig. 2 in Ref. \cite{sondenaa2006}). It is easy to see that C-type and G-type have $\hat{C}_4\hat{\Theta}$ symmetry combining $\hat{C}_4$ rotation along the z direction and TR operation. A large variety of materials with perovskite structures, such as SrMnO$_3$\cite{sondenaa2006}, PbFeO$_2$F\cite{katsumata2006}, etc, possess these symmetries. One needs to further search for semiconducting materials with a narrow gap in these systems in order to realize topologically non-trivial phases, which is beyond the scope of this paper.

\section{acknowledgments}
We would like to thank C. Fang, X.L. Qi and S.C. Zhang for useful discussions.

\appendix
\section{Review of co-representation theory for magnetic symmetry groups}
Now we consider the magnetic symmetry group ${\bf M}={\bf G}+A{\bf G}$ and assume the wave functions $|\psi_j\rangle$ ($j=1,\cdots,d$) form a set of basis for the unitary symmetry group ${\bf G}$. The corresponding representation is defined as $S|\psi_j\rangle=\sum_i\Delta_{ij}(S)|\psi_i\rangle$, which satisfies $\Delta(SW)=\Delta(S)\Delta(W)$ ($S,W\in {\bf G}$). Let's define $|\phi_j\rangle=A|\psi_j\rangle$, and
\begin{eqnarray}
	&&S|\phi_j\rangle=SA|\psi_j\rangle=A\sum_i\Delta_{ij}(A^{-1}SA)|\psi_i\rangle\nonumber\\
	&&=\sum_i\Delta^*_{ij}(A^{-1}SA)|\phi_i\rangle
	\label{eq:MSG_S1}
\end{eqnarray}
where we have used $A^{-1}SA\in{\bf G}$. For any $B\in A{\bf G}$, one has
\begin{eqnarray}
	&&B|\psi_j\rangle=A(A^{-1}B)|\psi_j\rangle=A\sum_i\Delta_{ij}(A^{-1}B)|\psi_i\rangle\nonumber\\
	&&=\sum_i\Delta^*_{ij}(A^{-1}B)|\phi_i\rangle
	\label{eq:MSG_B1}\\
	&&B|\phi_j\rangle=BA|\psi_j\rangle=\sum_i\Delta_{ij}(BA)|\psi_i\rangle
	\label{eq:MSG_B2}
\end{eqnarray}
Therefore $|\varphi\rangle$, defined as $|\varphi_{j}\rangle=|\psi_j\rangle$ and $|\varphi_{j+d}\rangle=|\phi_j\rangle$, form the basis for the symmetry group ${\bf M}$, and the corresponding ``representation'' $D$ is defined as
\begin{eqnarray}
	&&D(S)=\left(
	\begin{array}{cc}
		\Delta(S)&0\\
		0&\Delta^*(A^{-1}SA)
	\end{array}
	\right)	\label{eq:MSG_DS1}\\
	&&D(B)=\left(
	\begin{array}{cc}
		0&\Delta(BA)\\
		\Delta^*(A^{-1}B)&0
	\end{array}	\right)\label{eq:MSG_DB1}
\end{eqnarray}
where $S|\varphi_j\rangle=\sum_iD_{ij}(S)|\varphi_i\rangle$ and $B|\varphi_j\rangle=\sum_iD_{ij}(B)|\varphi_i\rangle$. We usually denote $\bar{\Delta}(S)=\Delta^*(A^{-1}SA)$ which also forms a representation for unitary group ${\bf G}$.
Although $D$ is consisted of two sets of representations, it is not the conventional representation by itself, which can easily be seen by multiplication rules.
\begin{eqnarray}
	&&D(BS)=\left(
	\begin{array}{cc}
		0&\Delta(BSA)\\
		\Delta^*(A^{-1}BS)&0
	\end{array}
	\right)\nonumber\\
	&&=\left(
	\begin{array}{cc}
		0&\Delta(BA)\Delta(A^{-1}SA)\\
		\Delta^*(A^{-1}B)\Delta^*(S)&0
	\end{array}
	\right)\nonumber\\
	&&=\left(
	\begin{array}{cc}
		0&\Delta(BA)\\
		\Delta^*(A^{-1}B)&0
	\end{array}
	\right)\left(
	\begin{array}{cc}
		\Delta^*(S)&0\\0&\Delta(A^{-1}SA)
	\end{array}
	\right)\nonumber\\
	&&=D(B)D^*(S)
	\label{eq:MSG_DBS1}
\end{eqnarray}
Therefore, $D$ does not obey the multiplication rule of the conventional representations, and it is usually called co-representation of magnetic symmetry group ${\bf M}$\cite{wigner1959,bradley1968}. Other multiplication rules can be obtained as
\begin{eqnarray}
	&&D(S)D(W)=D(SW)\label{eq:MSG_DSW1}\\
	&&D(S)D(B)=D(SB)\label{eq:MSG_DSB1}\\
	&&D(B)D^*(C)=D(BC)\label{eq:MSG_DBC1}
\end{eqnarray}

It is important to note that for a unitary transformation $U$, $|\varphi'_i\rangle=\sum_iU_{ji}|\varphi_j\rangle$, one has $S|\varphi'_i\rangle=\sum_j SU_{ji}|\varphi_j\rangle=\sum_{kj}U_{ji}D_{kj}(S)|\varphi_k\rangle=\sum_{kjl}U^{-1}_{lk}D_{kj}(S)U_{ji}|\varphi'_l\rangle$, and similarly $B|\varphi'_i\rangle=\sum_i BU_{ji}|\varphi_j\rangle=\sum_{jk}U^*_{ji}D_{kj}(B)|\varphi_k\rangle=\sum_{jkl}U^{-1}_{lk}D_{kj}(B)U^*_{ji}|\varphi'_l\rangle$,
so
\begin{eqnarray}
	D'(S)=U^{-1}D(S)U\\
        D'(B)=U^{-1}D(B)U^*.
	\label{eq:MSG_transformation}
\end{eqnarray}
From the above equaitons, it is easy to know that $Tr(D'(B))\neq Tr(D(B))$, so the character, as well as the character table, has no meaning for co-representation.


Although the character and character table can not work for co-representations, the general relation between degeneracy and an irreducible representation still exist. By construction, our co-representation $D$ has at least the dimension two. Therefore, one need to study the irreducibility of the co-representation $D$. It turns out that there is a simple rule to determine the reducibility of the co-representation $D$, known as the Herring rule\cite{herring1937a}. The Herring rule can be summarized as
\begin{eqnarray}
	\sum_{B\in A{\bf G}} \chi(B^2)=\left\{
	\begin{array}{c}
		|G|\qquad \mbox{in case a},\\
		-|G|\qquad \mbox{in case b},\\
		0\qquad \mbox{in case c}.
	\end{array}
	\right.
	\label{eq:MSG_Herring}
\end{eqnarray}
where $\chi$ is the character of the representation $\Delta$, $|G|$ is the element number of the group ${\bf G}$. The case a is reducible while the cases b and c are irreducible.

\section{Bloch Hamiltonian of $\hat{C}_4\hat{\Theta}$-invariant topological insulator}
In this section, we will present the detailed form of our our $\hat{C}_4\hat{\Theta}$-invariant tight-binding model in the momenmtum space. This Bloch Hamiltonian is given by a $2\times2$ block matrix.
\begin{equation}
H = \left(
\begin{array}{cc}
H_A & H_{AB} \\
H_{AB}^{\dagger} & H_B \\
\end{array}
\right)
\end{equation}
Each block $H_{A,B}$ is an $8\times8$ matrix, as is shown below.
\begin{widetext}
\begin{eqnarray}
& H_A^{11} & = \left(
\begin{smallmatrix}
 0 & -i M_1 &c_0 e^{ i b(k_y-k_x)}  (t_{p1}+ t_{s1}) & c_0 e^{ib (k_{y}- k_{x})} (t_{p1}- t_{s1}) \\
 i M_{1} & 0 & c_0 e^{ ib ( k_{y}- k_{x})} ( t_{p1}-t_{s1}) & c_0 e^{ib (k_{y}- k_{x})} ( t_{p1}+ t_{s1}) \\
 c_0 e^{ib (k_{x}- k_{y})} ( t_{p1}+ t_{s1}) & c_0 e^{ib (k_{x}- k_{y})} ( t_{p1}- t_{s1}) & 0 & i M_{1} \\
c_0 e^{ib (k_{x}- k_{y})} (t_{p1}- t_{s1}) & c_0 e^{ib i (k_{x}- k_{y})} ( t_{p1}+ t_{s1}) & -i M_{1} & 0 \\
\end{smallmatrix}
\right) \nonumber \\
& H_A^{22} & = \left(
\begin{smallmatrix}
 0 & -i M_1 &c_0 e^{i b (k_x-k_y)}  (t_{p1}+ t_{s1}) & c_0 e^{i b (k_{x}- k_{y})} (t_{p1}- t_{s1}) \\
 i M_{1} & 0 & c_0 e^{i b ( k_{x}- k_{y})} ( t_{p1}-t_{s1}) & c_0 e^{i b (k_{x}- k_{y})} ( t_{p1}+ t_{s1}) \\
 c_0 e^{i b (k_{y}- k_{x})} ( t_{p1}+ t_{s1}) & c_0 e^{i b (k_{y}- k_{x})} ( t_{p1}- t_{s1}) & 0 & i M_{1} \\
c_0 e^{i b (k_{y}- k_{x})} (t_{p1}- t_{s1}) & c_0 e^{i b (k_{y}- k_{x})} ( t_{p1}+ t_{s1}) & -i M_{1} & 0 \\
\end{smallmatrix}
\right)  \nonumber \\
& H_A^{12} & = \left(
\begin{smallmatrix}
  e^{-i2b k_x} t_{s1}+e^{i\bar{2b} k_x} t_{s3} & 0 &c_0 e^{ib (-k_x-k_y)}  (t_{p1}+ t_{s1}) & c_0 e^{ib (-k_{x}- k_{y})} (-t_{p1}+ t_{s1}) \\
0 & e^{-i2b k_x} t_{p1}+e^{i\bar{2b} k_x} t_{p3} & c_0 e^{ ib ( -k_{x}- k_{y})} (- t_{p1}+t_{s1}) & c_0 e^{ib (-k_{x}- k_{y})} ( t_{p1}+ t_{s1}) \\
 c_0 e^{ib (-k_{y}- k_{x})} ( t_{p1}+ t_{s1}) & c_0 e^{ib (-k_{y}- k_{x})} ( -t_{p1}+ t_{s1}) & e^{-i2b k_y} t_{p1}+e^{i\bar{2b} k_y} t_{p3} & 0 \\
c_0 e^{ib (-k_{y}- k_{x})} (-t_{p1}+ t_{s1}) & c_0 e^{ib (-k_{y}- k_{x})} ( t_{p1}+ t_{s1}) & 0 & e^{-i2b k_y} t_{s1}+e^{i\bar{2b} k_y} t_{s3} \\
\end{smallmatrix}
\right)  \nonumber  \\
& H_A^{21} & = (H_A^{12})^{\dagger}  \nonumber \\
& H_B & = H_A \text{  (hopping coefficients (including $M_1$) with an opposite sign)}
\nonumber \\
& H_{AB}^{11} &=
\left(
\begin{smallmatrix}
 e^{-ib k_{z}} t_{p2}+e^{i\bar{b} k_{z}} t_{p5} & 0 & a_1 & \bar{a}_1 \\
 0 & e^{-ib k_{z}} t_{p2}+e^{i\bar{b} k_{z}} t_{p5} & \bar{a}_1 & a_1 \\
a_2 & \bar{a}_2 & e^{-ib k_{z}} t_{p2}+e^{i \bar{b} k_{z}} t_{p5} & 0 \\
\bar{a}_2 & a_2 & 0 & e^{-ib k_{z}} t_{p2}+e^{i\bar{b} k_{z}} t_{p5} \\
\end{smallmatrix}
\right)
\nonumber \\
& H_{AB}^{22} & = H_{AB}^{11}(e_1\rightarrow e_2,e_2\rightarrow e_1)
\nonumber \\
& H_{AB}^{12} & =
\left(
\begin{array}{cccc}
a_{3} & 0 & a_5 & \bar{a}_5 \\
 0 & a_{4} & \bar{a}_5 & a_5 \\
 a_5 & \bar{a}_5 & \bar{a}_{4} & 0 \\
 \bar{a}_5 & a_5 & 0 & \bar{a}_{3} \\
\end{array}
\right)
\nonumber \\
& H_{AB}^{21} & = H_{AB}^{12}(k_x\rightarrow -k_x,k_y\rightarrow -k_y)
\end{eqnarray}
\end{widetext}

For each hopping amplitude, we have used lower indices s and p to label $\sigma$ and $\pi$ bond contribution. For example, $t_{s1}$ is the $\sigma$ contribution for $t_1$, while $t_{p1}$ is the $\pi$ contribution for $t_1$. We define $b=0.1$, $\bar{b}=1-b$ and $\bar{2b}=1-2b$. So $\sqrt{2}b$ is the distance between nearest neighboring atoms in one unit cell. We define $c_0=\frac{1}{2}$, $c_1=0.987952$, $c_2=0.984615$, $c_3=0.952941$, $c_4=\frac{1}{3}$ and $\bar{c}_i=1-c_i$. These $c_i$ and $\bar{c}_i$ are related to directional cosines of atom configurations. We also define $e_1=e^{ i b (-k_x+k_y-k_z)} $, $e_2=e^{i b (k_x-k_y-k_z)}$ , $e_3=e^{ i b (-k_x-k_y-k_z)}$,$\bar{e}_i=e_i e^{i k_z}$ and the notations below for simplicity. All the $a$ parameters are listed as follows.

\begin{widetext}
\begin{eqnarray}
a_1 & = & e_1 (\bar{c}_4 t_{p2}+c_4 t_{s2})+\bar{e}_1 (c_1 t_{p5}+\bar{c}_1 t_{s5}) \nonumber \\
\bar{a}_1 & = & e_1 (c_4 t_{p2}-c_4 t_{s2})+\bar{e}_1 (\bar{c}_1 t_{p5}-\bar{c}_1 t_{s5}) \nonumber \\
a_2 & = & e_2 (\bar{c}_4 t_{p2}+c_4 t_{s2})+\bar{e}_2 (c_1 t_{p5}+\bar{c}_1 t_{s5}) \nonumber \\
\bar{a}_2 & = & e_2 (c_4 t_{p2}-c_4 t_{s2})+\bar{e}_2 (\bar{c}_1 t_{p5}-\bar{c}_1 t_{s5}) \nonumber \\
a_{3} & = &  e^{i (-2b k_x-b k_z)}((2b t_{p2}+\bar{2b} t_{s2})+e^{i k_x} (\bar{c}_2 t_{p4}+c_2 t_{s4})+e^{i k_z} (c_3 t_{p5}+\bar{c}_3 t_{s5})) \nonumber \\
\bar{a}_{3} & = &  e^{i (-2b k_y-b k_z)}((2b t_{p2}+\bar{2b} t_{s2})+e^{i k_y} (\bar{c}_2 t_{p4}+c_2 t_{s4})+e^{i k_z} (c_3 t_{p5}+\bar{c}_3 t_{s5})) \nonumber \\
a_{4} & = & e^{i (-2b k_{x}-b k_{z})} t_{p2}+e^{i (\bar{2b} k_{x}-b k_{z})} t_{p4}+e^{i (\bar{b} k_{z}-2b k_{x})} t_{p5} \nonumber \\
\bar{a}_{4} & = & e^{i (-2b k_{y}-b k_{z})} t_{p2}+e^{i (\bar{2b} k_{y}-b k_{z})} t_{p4}+e^{i (\bar{b} k_{z}-2b k_{y})} t_{p5} \nonumber \\
a_{5} & = & e_3 (\bar{c}_4 t_{p2}+c_4 t_{s2})+\bar{e}_3 (c_1 t_{p5}+\bar{c}_1 t_{s5}) \nonumber \\
\bar{a}_5 & = & e_3 (c_4 t_{s2}-c_4 t_{p2})+\bar{e}_3 (\bar{c}_1 t_{s5}-\bar{c}_1 t_{p5})
\end{eqnarray}

\end{widetext}
%



\end{document}